\documentclass[doublespacing]{elsart}
\usepackage{natbib}
\usepackage{graphicx}
\usepackage{amssymb}

\newcommand{\Tr}{\mbox{Tr}}

\newcommand{\pdiffl}[2]{\frac{\partial #1}{\partial #2}}

\newcommand{\dfrac}[2]{\displaystyle\frac{#1}{#2}}
\newcommand{\grad}{\mbox{grad}}

\begin{document}

\begin{frontmatter}

\date{LA-UR-05-4739 -- May 31, 2005}

\title{Dynamic plasticity of beryllium \\ in the inertial fusion fuel capsule
   regime}
\author{Damian~C.~Swift\corauthref{dcs}},\ead{damian.swift@physics.org}
\corauth[dcs]{Corresponding author}
\author{Thomas~E.~Tierney},
\author{Sheng-Nian~Luo},
\author{Roberta~N.~Mulford},
\author{George~A.~Kyrala},
\author{Randall~P.~Johnson},
\author{James~A.~Cobble},
\author{David~L.~Tubbs},
\author{Nelson~M.~Hoffman}
\address{Los Alamos National Laboratory, \thanksref{doe}
   Los Alamos, NM 87545, USA}
\thanks[doe]{This work was performed under the auspices of the
            U.S. Department of Energy under contract W-7405-ENG-36.}

\begin{abstract}
The plastic response of beryllium was investigated on nanosecond time scales
appropriate for inertial confinement fusion fuel capsules,
using laser-induced shock waves, with the response probed with
surface velocimetry and in-situ x-ray
diffraction. Results from loading by thermal x-rays (hohlraum) were consistent
with more extensive studies using laser ablation. Strong elastic waves were
observed, up to $\sim$1 km/s in free surface speed, with significant structure
before the arrival of the plastic shock. The magnitude and shape of the
precursor could be reproduced with a plasticity model based on dislocation
dynamics. Changes in lattice spacing measured from the x-ray diffraction
pattern gave a direct measurement of uniaxial compression in the elastic wave,
triaxial flow from the decay of the precursor, and triaxial compression in the
plastic shock; these were consistent with the velocity data. The dynamic
strength behavior deduced from the laser experiments was used to help interpret
surface velocity data around the onset of shock-induced melting.  A model of
heterogeneous mixtures is being extended to treat anisotropic components,
and spall.
\end{abstract}


\end{frontmatter}




\section{Introduction}
The dynamic plasticity of beryllium on nanosecond time scales is important
in the development of inertial confinement fusion (ICF)
\citep{Wilson98,Bradley99,Lindl98}.
The initial experiments on thermonuclear ignition at the
National Ignition Facility will use 192 laser beams to deliver 1-2\,MJ
of energy inside a hohlraum.
The resulting flux of soft x-rays will implode a spherical capsule
containing cryogenic deuterium-tritium (DT) fuel.
If the implosion is sufficiently vigorous and uniform then thermonuclear
burn is predicted to propagate from the central hotspot.
The implosion has a relatively high aspect ratio
(initial to final radius) of around 30, and is unstable with respect to
spatial variations in thickness, surface finish, or radiation drive.

A leading material for the fuel capsule is Be containing up to 
around 1~weight percent of Cu to increase its opacity to the radiation drive. 
Compared to alternative capsule materials such as plastic,
Be is dense and has a high thermal conductivity, which helps maintain a
uniform layer of DT ice.
However, Be exhibits elastic and plastic anisotropy
when subjected respect to mechanical or thermal loading.
This anisotropic response is likely to induce velocity fluctuations from the
hohlraum drive, which may seed significant implosion instabilities.
It is thus necessary to define a specification for the microstructure
-- e.g. a maximum grain size -- for Be-Cu in an ICF capsule.
The response of Be on relevant length and time scales, 
$\sim$100\,$\mu$m and $\sim$10\,ns, has not been characterized previously,
so this work includes the development and calibration of appropriate models,
continuum mechanics simulations in which the microstructure is resolved
in order to predict the velocity fluctuations directly, 
and the experimental validation of predictions of anisotropic response and
instability seeding.
The continuum mechanics simulations are computationally expensive in three
dimensions, so some systematic studies of the velocity fluctuations are
performed in fewer dimensions.
A mixture model, originally developed for reactive flow studies 
\citep{Mulford01} but subsequently used for other applications 
including the shock compression of porous minerals \citep{Luo04_jpcm},
was generalized to predict the average response of strong, polycrystalline
materials, as an inline treatment of texture at coarser resolutions.

This paper discusses experiments performed to investigate the elastic-plastic
response of Be on nanosecond time scales, and the development of 
physically-based models to describe the behavior.
The mixture model, with extensions for strength, is presented.
The treatment of strength should allow material damage and failure to
be modeled seamlessly; damage pertinent to dynamic failure by spallation
is discussed.

\section{Ablative loading experiments}
\begin{figure}
\begin{center}
\includegraphics[scale=1.0]{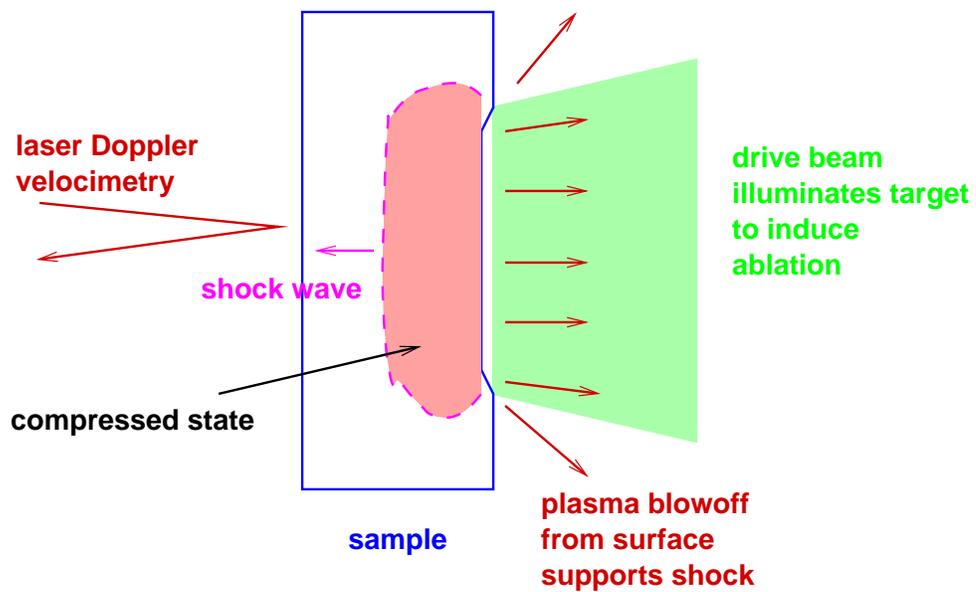}
\end{center}
\caption{Schematic cross-section of laser drive experiments.
   The sample was 10 to 50\,$\mu$m thick; the focal spot was
   5\,mm in diameter.}
\label{fig:exptschem}
\end{figure}

Plastic flow in Be under dynamic loading was investigated experimentally
using laser ablation to induce shock waves.
Experiments were performed at the TRIDENT laser facility at Los Alamos,
which has been used to induce shocks and quasi-isentropic compression waves
by ablation of a free surface \citep{Swift04_ddshock_pre,Swift05_lice_pre},
and also to induce loading and accelerate flyer plates by ablation through
a transparent substrate to act as confinement \citep{Swift05_flyer_rsi}.
TRIDENT has a Nd:glass laser system with a fundamental wavelength of 1054\,nm.
For more efficient coupling with matter, the pulse was frequency-doubled to
527\,nm (green).
Operating in nanosecond mode, the pulse comprised up to 13 elements 180\,ps long
for each of which the intensity could be set independently,
allowing the irradiance history to be controlled, with a total energy of up
to 250\,J.
In each experiment, the planar sample was irradiated on one side,
using a Fresnel zone plate was distribute the laser energy uniformly over
a spot 5\,mm in diameter.
(Fig.~\ref{fig:exptschem}.)

\begin{figure}
\begin{center}
\includegraphics[scale=1.0]{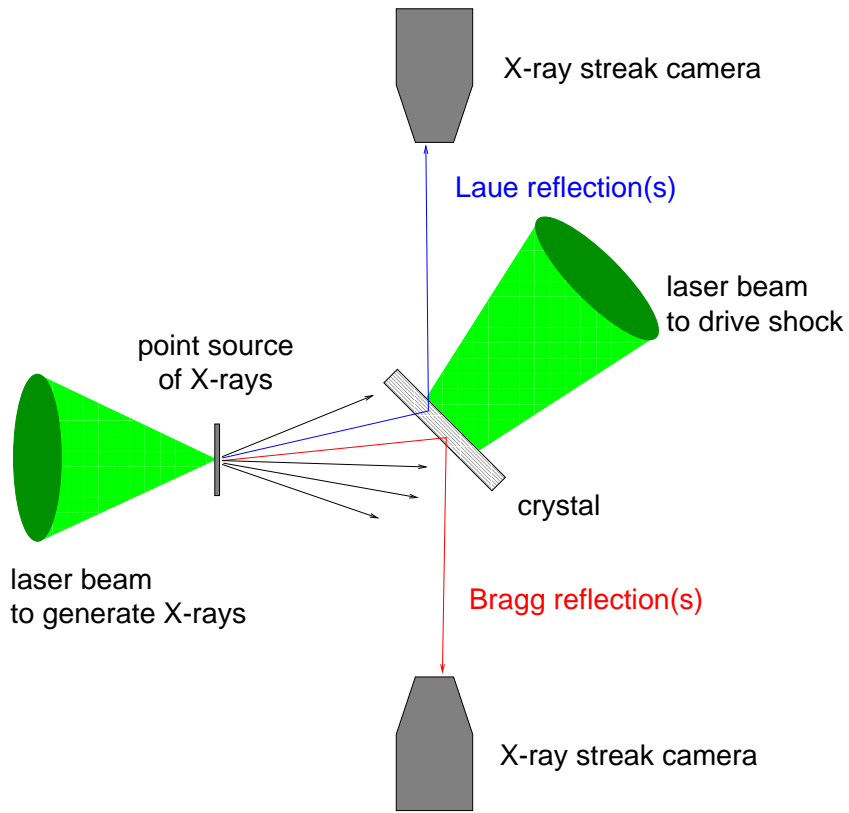}
\end{center}
\caption{Schematic of experimental configuration for transient
   x-ray diffraction measurements.
   The x-ray source was 10-15\,mm from the sample,
   and the snouts of the x-ray detectors were 50\,mm apart.}
\label{fig:txdschem}
\end{figure}

Velocity histories were obtained at the surface opposite the laser ablation,
by laser Doppler velocimetry over a line 600\,$\mu$m long centered opposite
the center of the ablation spot.
In some experiments, a second laser beam was used to generate x-rays
by focusing it over a $\sim$150\,$\mu$m spot on a Ti foil, 10\,$\mu$m thick;
the plasma resulting from a pulse energy of 100-200\,J in 1-2.5\,ns
cooled by emitting predominantly He-$\alpha$-like line radiation
(Fig.~\ref{fig:txdschem}).
The x-rays were used to obtain diffraction measurements {\it in situ}
during the shock wave event and thus to monitor
the response of the crystal lattice directly \citep{Swift01_sieos_prb}.

To simplify the analysis and interpretation of shock wave experiments,
it is desirable that the loading history should be constant for some period,
then fall rapidly to zero.
A disadvantage in using ablative loading compared with flyer impact
experiments is that the relation between irradiance history and pressure history
applied to the sample is not simple: for instance, a constant irradiance does 
not lead to a constant drive pressure \citep{Swift04_ddshock_pre}.
The irradiance history must in general be tailored to the material, desired
pressure, and pulse duration.
Often on nanosecond time scales, a constant pressure requires an irradiance
history which increases by several tens of percent over the duration of the
pulse.
Tailored pulse shapes such as this can be generated at TRIDENT because of
the flexibility in pulse construction, but would generally require several
iterations because of inaccuracies in radiation hydrodynamic simulations used
to predict the desired shape, making it necessary to perform experiments
on shock flatness to verify and adjust the shape.
The pulse shape and energy produced by the laser were not perfectly
reproducible, so there is a limit to the accuracy with which the pressure 
history can be specified -- e.g., constant -- on a given shot.
Rather than attempting to use an ideal irradiance history,
the pulse shape was measured on each shot and used in radiation hydrodynamics
simulations to predict the pressure history at a point close to the
ablated surface.
The pressure history was then used as a time-dependent boundary condition
in continuum mechanics simulations, allowing more detailed models of the
behavior of condensed matter to be used without requiring the simulation
program to include radiation physics also.
This procedure has been found to work well for a range of applications
on metallic and non-metallic elements, and intermetallic compounds
\citep{Swift04_ddshock_pre,Swift04_ddalloy_pre,Swift05_lice_pre},
largely because
relevant material properties in the plasma state -- the equation of state (EOS),
opacity, and conductivities -- can be predicted with adequate accuracy.
For the pressures of a few tens of gigapascals considered here,
the radiation hydrodynamics simulations predicted that 
less than a micrometer of material was ablated.
Recovered samples have shown only a thin ($\sim$1\,$\mu$m) layer of the 
remaining material significantly affected by the heat associated with the 
ablation process.

At a few tens of micrometers, the sample thickness was two orders of magnitude 
less than the spot size, so the experiments were one-dimensional during the 
passage of the initial shock and subsequent release from the drive surface and
the surface monitored by velocimetry.
One-dimensional simulations of radiation hydrodynamics and continuum mechanics
were adequate.
The radiation hydrodynamics simulations used the computer programs
`LASNEX' \citep{LASNEX} and `HYADES' \citep{HYADES}.
The continuum mechanics simulations used the program `LAGC1D' \citep{LAGC1D}.
These programs used second-order finite difference representations 
of space and time with Lagrangian (material-following) derivatives,
and artificial viscosity to stabilize shocks.

\section{Shock response data for Be}
\begin{figure}
\begin{center}
\includegraphics[scale=1.0]{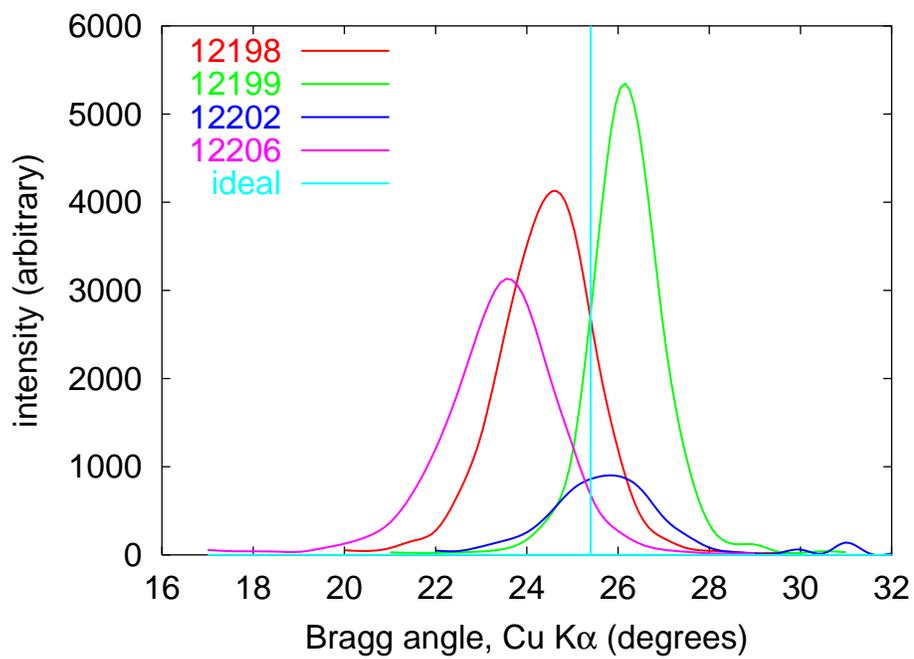}
\end{center}
\caption{Rocking curves (distribution of plane orientation with respect
   to surface) for nominally $(0001)$ beryllium crystals.
   Numbers are TRIDENT shot indices.}
\label{fig:xtalrocking}
\end{figure}

\begin{figure}
\begin{center}
\includegraphics[scale=0.8]{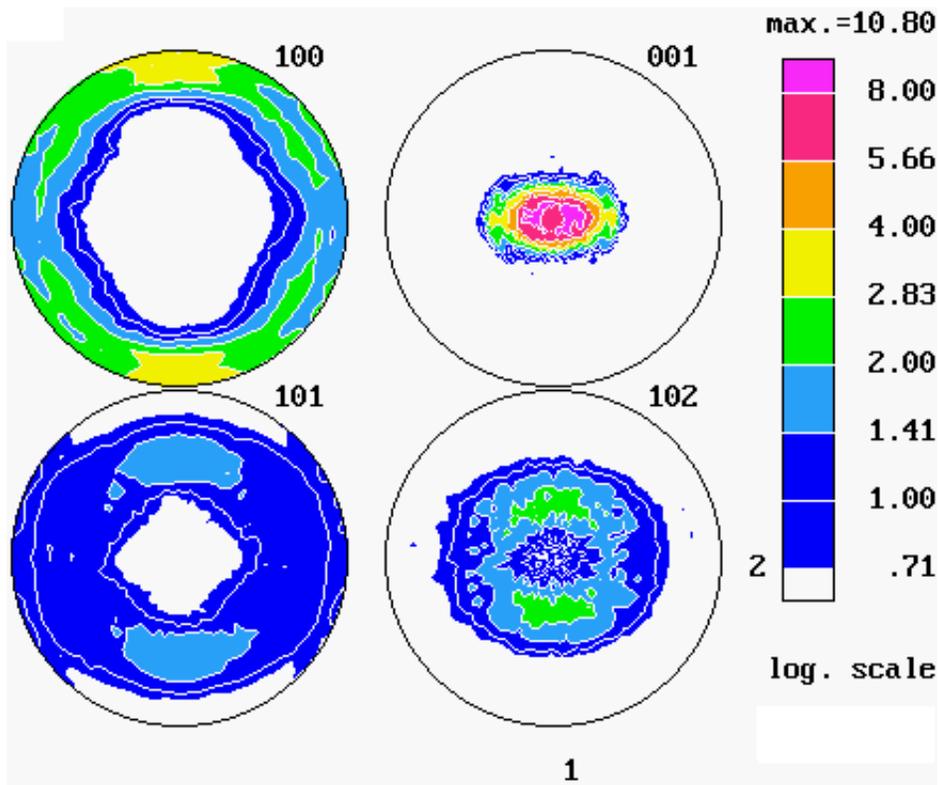}
\end{center}
\caption{Orientation imaging maps of rolled beryllium foil.
   Each circle shows the distribution of orientation of a particular
   Bragg reflection as a stereographic projection centered on the normal
   to the foil.
   The color map shows the density of the reflection at each angle,
   on a logarithmic scale.
   The plotting software used primitive plane indices, so `100'
   refers to the hexagonal plane $(10\bar 10)$ etc.}
\label{fig:foilorient}
\end{figure}

The plasticity experiments on Be included single crystals and rolled foils.
The crystals were cut nominally parallel to $(0001)$ planes 
from a boule grown by zone refining,
and were polished to 30-40\,$\mu$m thick.
The rocking curves were around $2^\circ$ wide, full width, half maximum
(Fig.~\ref{fig:xtalrocking}).
The rocking curve was determined statically by diffraction of Cu $K_\alpha$
x-rays, integrated over almost the whole of each sample.
The profile of the dynamic diffraction lines was consistent with
the static rocking curve, though there was some sign that
in the dynamic diffraction geometry, where the source was quite close to the 
sample, the width of the diffracted lines was slightly narrower.
The foils were commercial high-purity material (at least 99.8\%\ Be).
The texture, measured by x-ray orientation mapping, was found to be
with $c$-axes oriented predominantly normal to the foil -- i.e., similar to
the single crystals but with a distribution of angles $\sim 20^\circ$ wide
(full width) --
and the $a$-axes distributed continuously in the plane,
though clearly exhibiting preferential directions
(Fig.~\ref{fig:foilorient}).
Compared with the single crystals, the foils
presumably had been subjected to a much greater degree of plastic work.

\begin{figure}
\begin{center}
\includegraphics[scale=1.0]{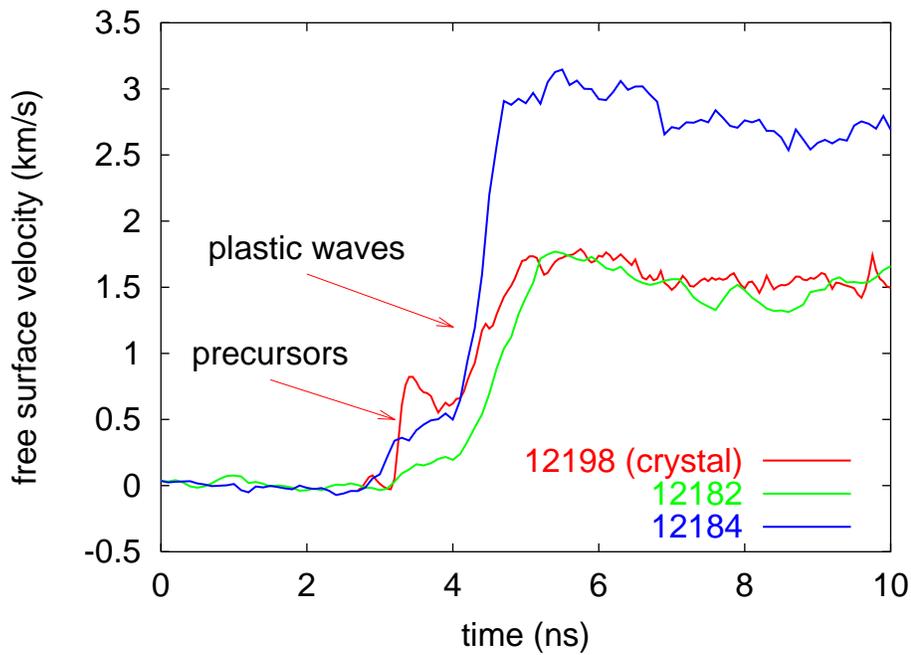}
\end{center}
\caption{Free surface velocity history from single crystal
   (TRIDENT shot 12198: 40\,$\mu$m $(0001)$, 2.7\,PW/m$^2$ for 1.8\,ns)
   and foils (shot 12182: 50\,$\mu$m, 3.2\,PW/m$^2$ for 1.8\,ns;
   shot 12184: 50\,$\mu$m, 6.5\,PW/m$^2$ for 1.8\,ns).
   Foil histories were displaced in time so the precursors occurred at
   a similar time to that in the crystal.}
\label{fig:uhist}
\end{figure}

\begin{figure}
\begin{center}
\includegraphics[scale=1.0]{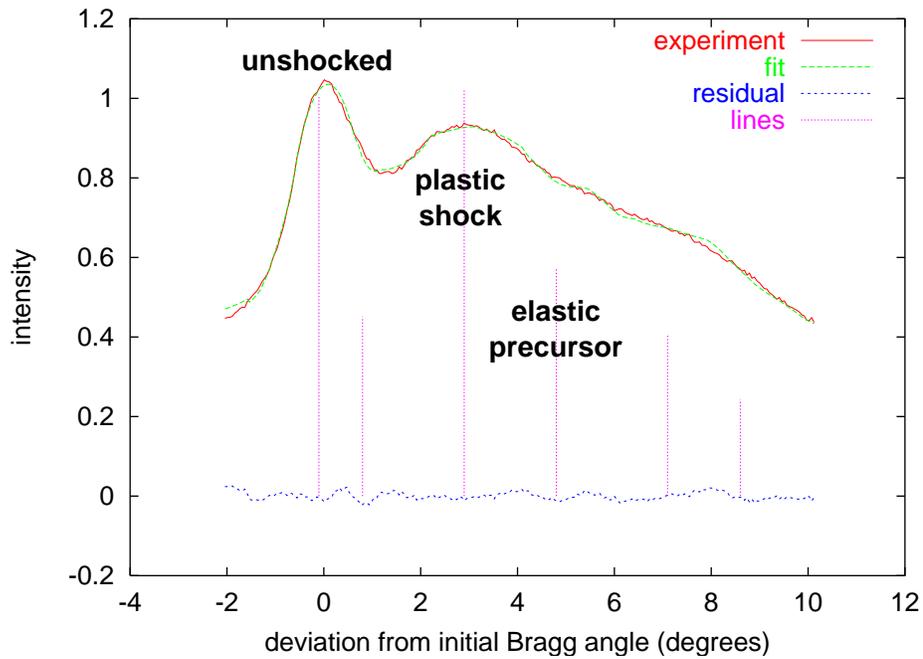}
\end{center}
\caption{Time-integrated x-ray diffraction data from single crystal
   and fit by deconvolving using measured rocking curve
   (TRIDENT shot 12202: 40\,$\mu$m $(0001)$, 6\,PW/m$^2$ for 1.8\,ns).
   Lines were identified by inspection and by comparison with compression
   predicted using radiation hydrodynamics simulations.}
\label{fig:txdunfold}
\end{figure}

\begin{figure}
\begin{center}
\includegraphics[scale=1.0]{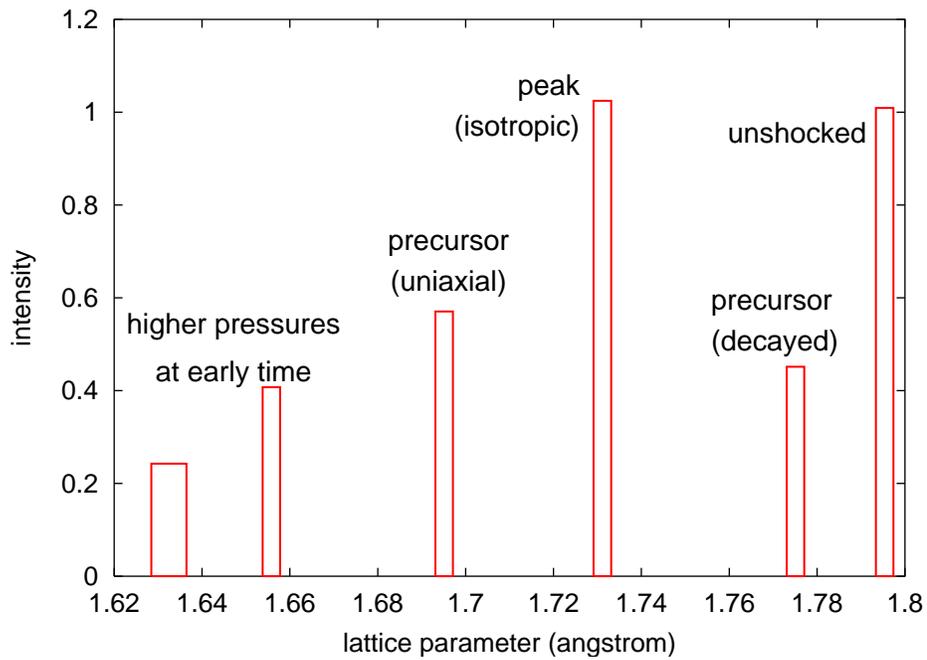}
\end{center}
\caption{Time-integrated x-ray diffraction data from single crystal,
   converted to variations in lattice parameter
   (TRIDENT shot 12202: 40\,$\mu$m $(0001)$, 6\,PW/m$^2$ for 1.8\,ns).
   Width of each bar indicates uncertainty.}
\label{fig:txdlatt}
\end{figure}

In the velocity history records from the single crystal experiments, 
the elastic precursor was very clear
with the lowest pressure drive (mean pressure $15.0$\,GPa).
The velocity showed a sharp initial peak (rise time faster than could be
resolved), followed by a deceleration,
followed by the main plastic wave (Fig.~\ref{fig:uhist}).
The x-ray diffraction records for the three lowest pressures
exhibited a peak corresponding to the uniaxial compression in the precursor
deduced from the initial peak, and at the lowest two pressures
an additional peak corresponding to the same volumetric compression
but expressed isotropically rather than uniaxially
-- suggesting that plastic relaxation to an isotropic state occurred
before the plastic wave arrived 
(Figs~\ref{fig:txdunfold} and \ref{fig:txdlatt}).
Velocity histories from experiments on rolled foils of similar thickness
showed a precursor of lower amplitude, with a finite and reproducible
rise time of around 400\,ps (Fig.~\ref{fig:uhist}).
The amplitude of the precursor varied more between experiments,
with higher amplitudes observed in thinner foils.
The higher amplitudes could be caused by time-dependence of plastic flow,
or by greater plastic strain imparted by rolling to a thinner foil.

\begin{figure}
\begin{center}
\includegraphics[scale=1.0]{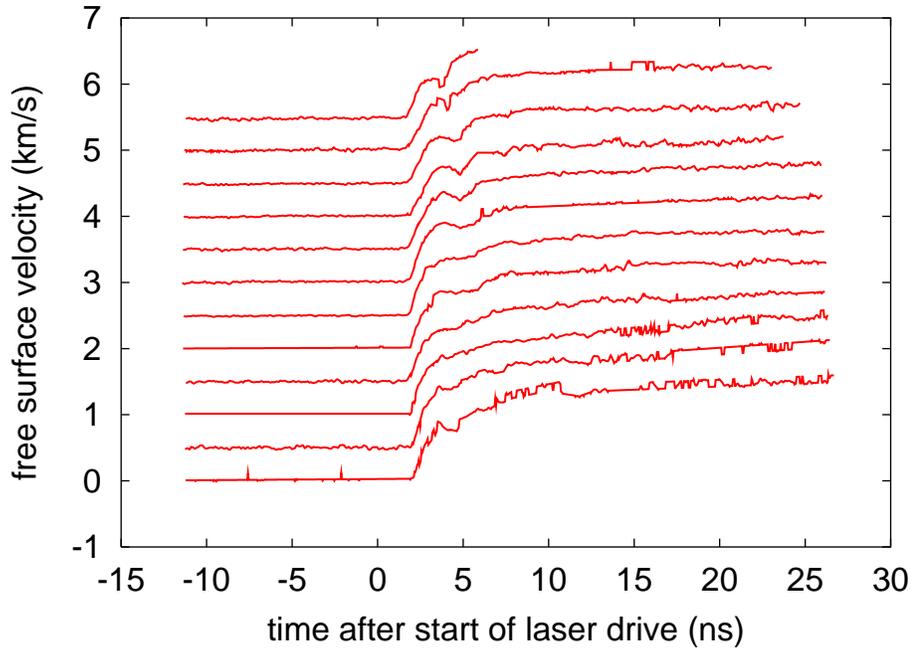}
\end{center}
\caption{Free surface velocity history from a beryllium foil,
   with traces taken at different points across the sample
   -- field of view $\sim 600$\,$\mu$m --
   showing spatial variations in the precursor
   (TRIDENT shot 12145: 13\,$\mu$m foil, 0.6\,PW/m$^2$ for 3.6\,ns).
   This experiment was driven at a relatively low pressure ($\sim 7$\,GPa),
   so the plastic wave was only slightly larger than the precursor;
   the precursor was also stronger than in the other foils, presumably
   because of greater cold working and/or time-dependence of flow.}
\label{fig:uxhist}
\end{figure}

The laser Doppler diagnostic recorded the velocity history over a line
across each sample. 
The records obtained from single crystals were spatially uniform,
demonstrating that the laser drive was adequately smooth over the
surface of the sample.
Some of the foil records exhibited spatial variations
in velocity following the precursor, in particular regions in which
there was a deceleration after the initial elastic peak,
resembling the single crystal velocity history
(Fig.~\ref{fig:uxhist}).

These velocity data -- particularly the single crystal record -- were compared
against simulations using different models of dynamic plasticity.
There was some uncertainty in sample thickness, and there was also some
timing jitter in the signal used to trigger the diagnostic cameras, so
timing variations of up to $\sim 1$\,ns were plausible.

\section{Simulations with different plasticity models}
Several models have been developed to describe the plastic response of Be 
under dynamic loading \citep{Steinberg96},
based on gas gun experiments using samples of the order of millimeters thick.
The laser-driven samples were a few tens of micrometers thick, so 
any time-dependence in the plastic flow processes would be expected to lead
to a higher flow stress and thus a higher velocity in the elastic precursor;
this is in fact what was observed.
One of the published models (Steinberg-Lund) is time-dependent, 
but these experiments explored time scales far shorter than the regime in which 
the model was calibrated, so one would not expect it necessarily to perform
well.

The speed of the free surface from the elastic precursor was a significant
fraction of the speed induced by the plastic shock.
The elastic wave reaches the surface first, reflects from it, and
rattles between the surface and the shock.
For a given shock pressure, the maximum speed seen at the free surface
depends on the amplitude of the elastic wave.
For a perfect model of the drive pressure, one would not expect the
peak free surface velocity to match the experiment unless the elastic
wave was modeled adequately.

Spall was ignored in most of the simulations, 
so the calculated surface velocity decelerated
close to zero after the end of the drive pulse.

\subsection{Hydrodynamic shock}
\begin{figure}
\begin{center}
\includegraphics[scale=1.0]{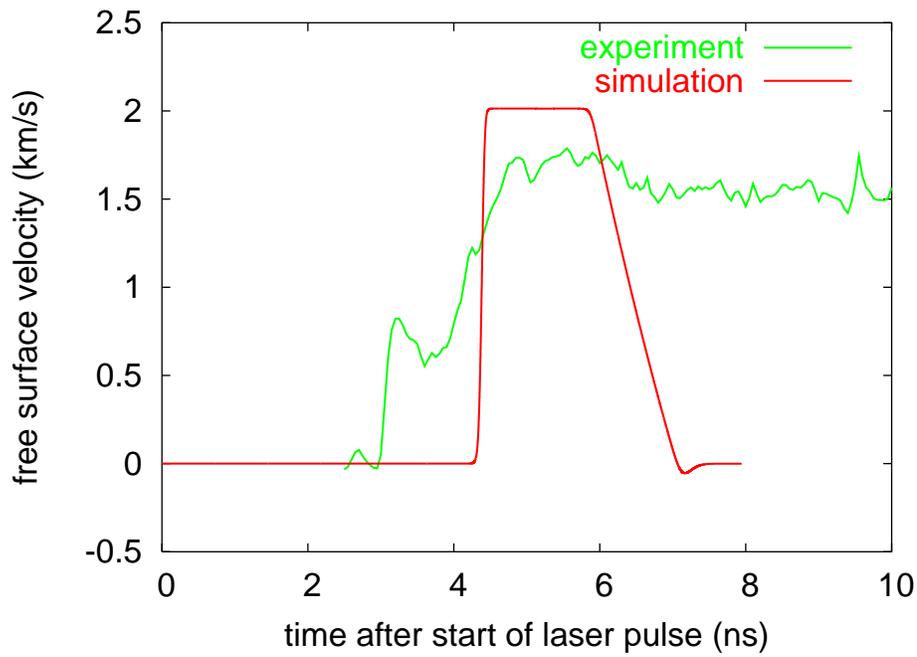}
\end{center}
\caption{Comparison between free surface velocity history measured for
   ablative loading of a single crystal cut parallel to $(0001)$
   (TRIDENT shot 12198: 40\,$\mu$m $(0001)$, 2.7\,PW/m$^2$ for 1.8\,ns)
   and a hydrodynamic simulation in which spall was ignored.}
\label{fig:hydro}
\end{figure}

For reference, simulations were performed with no strength model,
i.e. the scalar EOS only.
A cubic Gr\"uneisen form was used \citep{Steinberg96};
it was found previously that all common EOS agree for the mechanical response
of Be in the regime explored by these experiments.
With no elastic wave, the peak surface velocity was higher than
observed;
the arrival time of the shock was consistent with the experiment.
(Fig.~\ref{fig:hydro}.)

\subsection{Elastic-perfectly plastic}
As the plastic behavior of Be had not previously been explored in the
time and length scales investigated here, it is prudent to perform
simulations using a model with as few parameters as possible.
In the elastic-perfectly plastic strength model,
the material state is described by the isotropic state
and the elastic strain deviator.
The constitutive model comprises the EOS, a shear modulus $G$,
and a flow (or yield) stress $Y$.
The elastic response is $\sigma = G\epsilon_e$;
if the norm of the stress 
$||\sigma|| < Y$ then $\dot e_p = 0$; otherwise $\dot e_p = \dot e_a$.

\begin{figure}
\begin{center}
\includegraphics[scale=1.0]{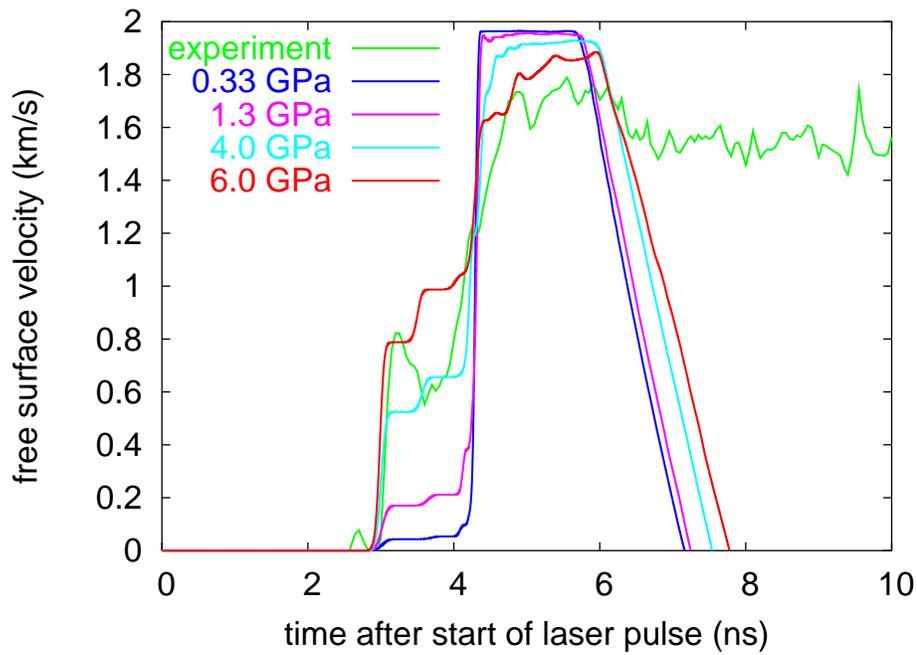}
\end{center}
\caption{Comparison between free surface velocity history measured for
   ablative loading of a single crystal cut parallel to $(0001)$
   (TRIDENT shot 12198: 40\,$\mu$m $(0001)$, 2.7\,PW/m$^2$ for 1.8\,ns)
   and elastic-perfectly plastic simulations using different values of
   the flow stress $Y$.}
\label{fig:ep}
\end{figure}

Parameters reported for beryllium are $G=151$\,GPa, and $Y$ between
0.33 and 1.3\,GPa.
With values in this range, the amplitude of the precursor was much lower
than observed experimentally.
The initial peak could be reproduced with $Y\simeq 6.0$\,GPa
and the following trough with $4.0$\,GPa;
the shape was not reproduced using this model.
With high values of the flow stress, the effect of the reverberating elastic
wave was evident as a sequence of steps in the acceleration history.
The steps were similar to some features seen in the experimental velocity
history.
As the flow stress was increased,
the predicted peak velocity decreased to approach the measured value,
and the release time increased to be close to the observation.
(Fig.~\ref{fig:ep}.)

\subsection{Steinberg-Guinan and Steinberg-Lund}
The Steinberg-Guinan model \citep{Steinberg96} is a generalization of the
elastic-plastic model to include work-hardening and thermal softening.
and varying the initial plastic strain $\epsilon_{p0}$ 
to investigate the sensitivity.
No set of parameters was found which was capable of reproducing the
detailed structure of the precursor.
If the maximum flow stress was increased to give an approximate match to the
amplitude of the precursor, the amplitude and duration of the plastic peak 
became very different to the observed velocity record.

The Steinberg-Lund model \citep{Steinberg96} is a generalization of the
Steinberg-Guinan model to include time-dependence.
Using the published parameters for Be, 
the amplitude of the elastic wave was lower than observed
and the plastic shock released much earlier then observed.
No set of parameters was found which reproduced the observed velocity history.

\subsection{Simplified microstructural plasticity}
Rather than attempting to test and calibrate explicit microstructural
plasticity models directly against the experimental data, 
simple extensions of the elastic-perfectly plastic model were made to
include time dependence with as few other parameters as possible.
Thus the systematic behavior observed in the single crystal experiments
could be investigated in isolation from the effects
of different terms in the more complicated models.

The plastic strain rate was chosen initially to be
\begin{equation}
\dot\epsilon_p = -\alpha G\epsilon\left[1 - \exp(-T^\ast/T)\right]
\end{equation}
where the shear modulus $G$, activation temperature $T^\ast$,
and rate multiplier $\alpha$ are 
in principle functions of the mass density $\rho$,
but were kept constant for the low compressions studied here.
The activation temperature was initially chosen to match the
Steinberg-Lund value, $T^\ast = 3600$\,K, and the sensitivity to
$\alpha$ and $T^\ast$ was explored.
It was possible to reproduce the observed wave profile qualitatively;
in particular, deceleration could be induced following the initial
elastic peak.
However, it was not possible to reproduce the precursor and the rise of the
plastic wave very accurately.
The simulations exhibited many small-amplitude spikes, caused by under-damped
reverberations of the precursor.
Work-hardening from plastic flow behind the initial wave would cause
a smearing of the plastic wave and should act to damp out the reverberations, 
but this was not investigated in the context of this model.

A slightly more realistic flow model was investigated to add the flexibility
to reproduce the characteristic decay time of the precursor
independently of its initial amplitude.
In reality, the barrier for defect hopping mechanisms in a crystal
is altered by the strain.
The potential of an atom or defect with respect to displacement is
periodic, with an amplitude like $\sin^2(\pi\epsilon/b)$ where 
$\epsilon$ is the linear displacement and $b$ the length of the
Burgers vector in the direction of interest.
Thermally-activated hopping may in principle occur in either the direction of
increasing or decreasing strain energy; the net rate is 
\begin{equation}
\dot\epsilon_p = \alpha(\rho) \left\{
   \exp[-T^\ast(1-\gamma)/T]-\exp[-T^\ast(1+\gamma)/T]\right\}
\end{equation}
where 
\begin{equation}
   \gamma(\epsilon_e)\equiv\sin^2\left(\frac{\pi\epsilon_e}{b(\rho)}\right).
\end{equation}
The reference period for elastic strain was taken to be a constant, 
$b(\rho)=1$.
The activation temperature was again chosen to match the
Steinberg-Lund value, $T^\ast = 3600$\,K, and the sensitivity to
$\alpha$ was explored.
For $\alpha\le 10$/ns, the material propagated an elastic wave.
At $\alpha=100$/ns, the shock had started to split into a precursor 
and a plastic wave, though elastic reverberations in the plastic wave were
pronounced.
For $\alpha\ge 1000$/ns, the precursor had the same peak-and-dip shape
as seen experimentally, with a step just less than halfway up the plastic
wave.
A good fit to the initial peak and the subsequent deceleration was obtained
for $\alpha=2000$/ns, though the acceleration during the plastic wave
was sharper.
As with the plastic relaxation model, overall rise time of the plastic wave
matched the experiment because of the delay introduced by the step
during the rise.
The release wave after the peak occurred at the same time,
and some structures in the peak resembled the experiment.

For defect-induced plastic strain, the rate is
$$
\dot\epsilon = R_d|\vec b|^2l
$$
where $l$ is the length of dislocations per volume of material,
$|\vec b|$ the Burgers vector (projected onto the direction of interest),
and $R_d$ the jump rate of a defect.
One could
assume that a defect jog must comprise $N$ atoms in order to be stable,
where $N$ could be calculated given the line energy of the defect compared
with the strain energy reduced by motion.
If the probability rate of a single atom jumping is $R_a$,
the probability rate of a stable jog forming is $R_a^N$.
As a simple estimate, the local jump rate $R_a$ occurs over a length $|\vec b|$,
so
$$
\dot\epsilon = R_a|\vec b|l.
$$
For Be, $|\vec b|\simeq 2$\,\AA.
The dislocation density can be estimated from the rocking curve, width $\theta$.
Assuming the curvature is caused by a population of defects of the same
orientation, the linear spacing between defects is $|\vec b|/\theta$,
so the defect density 
$$
l = \frac\theta{|\vec b|}.
$$
For a $2^\circ$ rocking curve in the Be crystals considered here, this implies
$l\simeq 1.7\times 10^{10}$/m. 
The rate at which individual atoms attempt to jump and thus mediate motion
of a defect is $\sim 10^{12}$\,Hz, thus the 
prefactor $\alpha=2000$/ns is consistent with the rate of deformation that
one would expect.

\begin{figure}
\begin{center}
\includegraphics[scale=1.0]{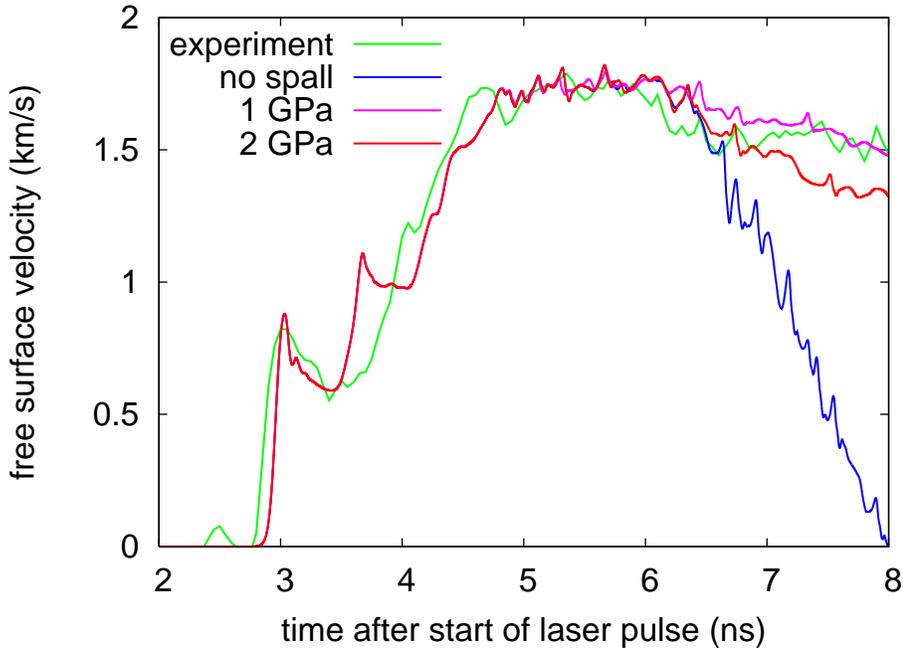}
\end{center}
\caption{Comparison between free surface velocity history measured for
   ablative loading of a single crystal cut parallel to $(0001)$
   (TRIDENT shot 12198: 40\,$\mu$m $(0001)$, 2.7\,PW/m$^2$ for 1.8\,ns)
   and simulations using the crystal relaxation model,
   ($T^\ast=3600$\,K and $\alpha=2000$/ns),
   with a simple spall model, for different values of the spall strength.}
\label{fig:pr2a_spall}
\end{figure}

Spallation of the sample should really be included through a failure model
intimately connected with the plasticity model. Here, simple spall
simulations were performed to test whether the results would be at all
consistent with the experiment, given the strong and structured elastic 
precursor.
Spall was modeled using the `minimum pressure' method: the EOS was modified
so that the minimum pressure it could produce was limited to a specified
value -- the spall strength, defined as a negative number -- and any greater
degree of tension would cause unbounded expansion, leading to a region of
low density.
The degree of deceleration after the plastic peak was reproduced with 
a spall stress $\sim$2\,GPa,
compared with a reported value from gas gun experiments of 0.35\,GPa
for Be alloy S200 \citep{Steinberg96}.
The difference is reasonable on grounds of rate-dependence.
(Fig.~\ref{fig:pr2a_spall}.)

\subsection{Microstructural plasticity}
The simple models discussed above are adequate to demonstrate that
microstructural processes explain the plastic behavior of a single
orientation of Be crystals, but a more complete model of rate-dependent
crystal plasticity is necessary for the resolved-microstructure simulations
of instability seeding for ICF; the discussion below follows the spirit of
\citet{Rashid92}.
The crystal lattice is described locally by the lattice vectors,
$\{a,b,c\}$; these can be defined with respect to standard conventions
of orientation ($a$ parallel to the $x$-axis, $b$ in the $xy$ plane) 
by a local rotation matrix, $R$.
The EOS and elasticity give the stress tensor $s(a,b,c)$.
Plastic flow is mediated by defects -- dislocations or disclinations --
which act along a set of slip systems defined with respect to the lattice
vectors.

Each slip system is characterized by a plane (normal $\vec n$) and 
a direction $\vec d$ in the plane.
Given the stress tensor $s$, the stress resolved along this slip system
is 
\begin{equation}
s_r = \vec d.s.\vec n = \vec d.\sigma.\vec n
\end{equation}
since $\vec n$ and $\vec d$ are orthogonal.
The plastic flow rate in the system is
\begin{equation}
\dot\epsilon_r = Z(\rho,T)\left\{
   \exp\left[-T^\ast(1-\gamma)/T\right]-\exp\left[-T^\ast(1+\gamma)/T\right]
   \right\}f(\rho_d/\rho_o)
\end{equation}
where $\rho_d$ is the dislocation density, $\rho_o$ the obstacle density,
$T^\ast$ the barrier energy (again expressed as a temperature),
and
\begin{equation}
\gamma(\epsilon_r)\equiv\sin^2\left(\frac{\pi\epsilon_e}{b(\rho)}\right)
\end{equation}
where 
\begin{equation}
\epsilon_r = \vec d.e.\vec n = \vec d.\epsilon.\vec n.
\end{equation}
Given a plastic strain rate $\dot\epsilon_r$ in a system, the strain tensor
changes as
\begin{equation}
\dot e = \dot\epsilon = -\sum_r\dot\epsilon_r\vec d\otimes\vec n
\end{equation}
\citep{Rashid92}.
Note that $\epsilon$ is always symmetric, though contributions from individual
slip systems may not be.
So far, the strain rate in a given slip system is identical with the 
simple hopping-based model above except for $O(1)$ 
geometrical terms describing the resolved shear stress on the system
and the resolved strain, and the explicit magnitude in terms of density
of dislocation length.
Further elastic precursor measurements on Be crystals of different orientations
will be needed for a complete calibration of the crystal plasticity model.

Work-hardening enters crystal plasticity through the interactions between
dislocations or twin boundaries and obstacles, including other dislocations.
The dislocation density and obstacle density increase with plastic strain:
\begin{equation}
\dot\rho_d = F_1(\rho,\rho_o)\dot\epsilon_r + F_2(\sigma_r),\quad
\dot\rho_o = G(\rho,\rho_o)\dot\epsilon_r
\end{equation}
where $F_1$ and $F_2$ are dislocation generation by flow 
(e.g. Frank-Reade sources) and by direct nucleation respectively.
These parameters will be calibrated against molecular dynamics simulations,
and validated by comparison with the slope and detailed shape on the
main plastic wave.

\section{Strength of heterogeneous mixtures}
A large amount of work has been done on interpreting shock wave data 
on heterogeneous materials -- particularly porous ones -- in terms of
properties of pure materials \citep{Trunin98}.
Work has also been done to predict restricted sets of mechanical properties
such as the shock Hugoniot, given the composition or initial porosity
\citep{Trunin98,Bushman93}.
However, relatively little has been done to develop models capable of
predicting wide-ranging response in a form satisfying the requirements
of general continuum mechanics simulations.

It is highly desirable to have a model which predicts the response of a
heterogeneous material given models for homogeneous components and a
representation of the microstructure.
For mixtures and materials of low porosity, 
models of the response for use in continuum simulations have been developed
from experiments on the heterogeneous material itself, without including a model
of the ``pure'' components
\citep{Asfanasenkov70,Thouvenin65,Meyers94}.
This approach is pragmatic, but has significant disadvantages.
If interest in a different composition grows, 
further experiments are needed to calibrate a new model.
{\it Ab initio} techniques are increasingly capable of predicting the response
of pure elements and of stoichiometric compounds; these theoretical models
cannot be applied to mixtures and porous materials without a predictive model
of heterogeneous materials.
Finally, for some applications, the composition or microstructure of a
heterogeneous material may vary continuously in space, or may evolve in time.
This is the case for sputtered Be-Cu alloy in ICF, where the texture may
change radially through the capsule, particularly in the presence of an ion
current.
A predictive, inline mixture model is desirable here for simulations
at varying degrees of detail from resolved microstructure to continuum-averaged
properties.

We have previously developed explicit mixture models for chemical explosives,
allowing the shock initiation properties to be predicted as a function of
composition and porosity
\citep{Swift99,Mulford01}.
These models were essentially hydrodynamic, i.e. they did not incorporate
elastic and plastic effects explicitly in the mixture model or the stress
field generated by the material.
Plasticity was included as a contribution to heating in the treatment of
hotspots.
Here we present a more general model in which the constitutive behavior of
the components is also included.

The dynamic response of a heterogeneous mixture is determined by the properties
of its components and the equilibration of stress and temperature between them.
For the propagation of a single shock wave of constant pressure,
the conditions of interest may be perfect equilibration.
However, there is no such thing as an `ideal' shock wave experiment:
in practice, experiments and real applications explore shock propagation 
through a finite thickness of material and hence probe a finite time scale.
If a steady shock is shown to exist, then the Rankine-Hugoniot equations
apply and equilibrium conditions apply behind the shock.
In practice, the steadiness is established with a finite accuracy,
and in general there may be equilibration processes occurring on a longer
time scale which are in effect frozen during the shock wave experiment.
Thus it is not always correct to assume that perfect equilibration occurs
across a shock wave.
In the case of more general loading histories, the response of the material
may depend on its instantaneous non-equilibrium state, so this should be 
taken into account.
It has been found by comparison with shock experiments on mixtures that 
mixture models which ignore equilibration and strength
effects are not able to reproduce all the observed shock behavior
\citep{Trunin98}.
Thermally-activated processes such as chemical reactions, plastic flow,
and phase transitions depend more strongly on non-equilibrium effects.

\begin{figure}
\begin{center}\includegraphics[scale=1.0]{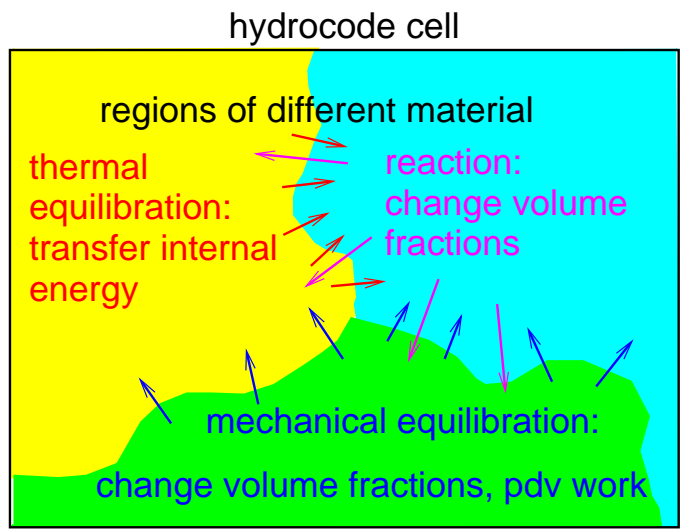}\end{center}
\caption{Schematic of the heterogeneous model.}
\label{fig:mixture}
\end{figure}

The model described below was developed to act as a bridge between
simulations with a resolved microstructure and continuum-averaged
simulations.
It is based on continuum-level averaging, but includes microstructural
terms which can be estimated fairly easily, or calibrated against
more detailed simulations.
Each component is described as a pure material, i.e. with its own
EOS and strength model.
The model was designed for use in continuum mechanics,
so it consists of a local state (varying with location and time
inside a material subjected to dynamic loading)
and a description of the material response (the same throughout
a particular material).
Since each component has its own state, a stress and temperature can be
calculated for each.
The material response model uses the local state to calculate
`external' properties such as mass density, stress and temperature,
and to calculate the evolution of the state given applied loading and
heating conditions.
Stress and temperature are equilibrated explicitly
according to a separate time constant for each.  This produces an
exponential approach to mechanical and thermal equilibrium
separately.  Ideal equilibration can be enforced by setting the time
constants to a small value.
Stress equilibration was performed by adjusting the strain state of
each component towards equilibrium.
In the absence of constitutive effects, stress equilibration 
becomes pressure equilibration;
this was performed by adjusting the volume fraction of
each component towards equilibrium, and expanding or compressing
along an isentrope.
Temperature equilibration was performed by transferring heat energy
at constant volume.
Equilibration based on finite rates was found to be convenient 
because it can be designed to use essentially the same material properties
as a continuum simulation of a `pure' material,
thus is was possible to implement the heterogeneous mixture model as 
a superset of existing types of material property (EOS etc) without
having to add additional types of calculation
(e.g. different derivatives of the EOS).
The numerical schemes are described in greater detail below.
The time scales for equilibration may be estimated 
from the grain size and properties of each component.
In principle, they could be considered as
additional continuum variables and evolved according to other
microstructural processes.
(Fig.~\ref{fig:mixture}.)

Porous materials were
represented by starting with a non-zero volume fraction of
gas (air or reaction products).  

\subsection{Representation of the microstructure}
We have previously reported results from reactive flow models using
mixtures of EOS 
\citep{Swift99,Mulford99b,Mulford01}.
These models treated
the mixture as a set of discrete components $i$ of volume fraction
$f_i$, each with its own state $s_i$.
This approach can be extended in a straightforward way
to equilibrate stresses rather than pressures.
The representation of the microstructure was contained entirely in
the volume fractions.
Here we describe an incremental improvement describing the microstructure
with slightly more detail.

\begin{figure}
\begin{center}\includegraphics[scale=1.0]{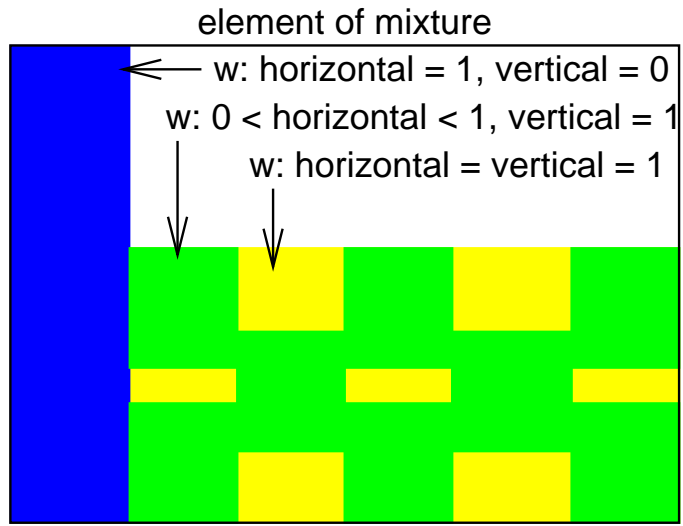}\end{center}
\caption{Schematic of the area overlap model.
   Outer rectangle represents a spatial domain described by the mixture model,
   e.g. a hydrocode cell.
   Different components of the mixture are represented by rectangles of
   different color.
   Overlap tensor $\omega$ describes the extent to which each component
   spans the whole domain, in a given direction.}
\label{fig:hetmix}
\end{figure}

For each component, a `fractional overlap area' $\omega_i$ was also defined;
this tensor quantity expresses the connectivity of the component across 
the mixture in each direction, in the same sense as stress and strain tensors.
The overlap areas were used in equilibrating stress, allowing a strong component
deemed to be continuous across the mixture to support a stress 
in the presence of
another component -- such as a void -- which could accommodate strain with
a smaller stress (Fig.~\ref{fig:hetmix}).
An overlap area of zero for a component in some orientation implies that the
component is continuous across the mixture in that orientation.
An overlap area of unity implies that the component stress should be 
equilibrated ideally in that orientation.
Intermediate values were used to simulate granular mixtures.

Each component in general has different material properties.
In addition to the properties pertaining to each `pure' component,
each component was associated with a tensor evolution parameter 
$d\omega/d\epsilon$ describing the rate of change of the overlap area with
plastic strain normal to the overlap orientation.
This parameter was used typically to simulate the homogenization of the mixture
as shear strain breaks the continuity of components across the mixture.
The evolution of the overlap areas was approximated by a material constant
(vector) $d\omega/d\epsilon$ where $\epsilon$ is interpreted as 
the strain component normal to the component of $\omega$.
A stress tensor and temperature is calculated for each component,
and for the mixture as a whole.

\subsection{Thermal equilibration}
Thermal equilibration acts to reduce differences between the temperature
$T_i$ in each component, which is a function of the instantaneous state $s_i$ 
in that component.
Given the component temperatures, we define a mean temperature for the
mixture as a whole, weighting by the heat capacity of each component:
\begin{equation}
\bar T \equiv \dfrac{w_i T_i}{\sum_i w_i}
\quad : \quad
w_i \equiv f_i c_v(s_i)\rho(s_i),
\end{equation}
where $f_i$ is the volume fraction of the $i$th component,
$c_v$ is its specific heat capacity, and $\rho$ its mass density.
Given the thermal relaxation time scale $\tau_T$, the relaxation factor 
over a time increment $\Delta t$ is
\begin{equation}
\phi_r \equiv \left(1 - e^{-\Delta t / \tau_T}\right).
\end{equation}
Thus the change in specific internal energy for each component can be
calculated for a finite time increment,
\begin{equation}
\Delta e_i = -\phi_r c_v(s_i) \left(T_i - \bar T\right).
\end{equation}

\subsection{Mechanical equilibration}
Mechanical equilibration acts to reduce differences between the stress
$\tau_i$ in each component, 
which is a function of the instantaneous state $s_i$ 
in that component, which includes the strain.
Here, stress and strain are tensor quantities, expressed in Voigt (vector)
notation if desired.
Given the component stresses, we define a mean stress for the
mixture as a whole:
\begin{equation}
\bar\tau \equiv f_i \tau_i(s_i)
\end{equation}
Given the mechanical relaxation time scale $\tau_m$, the relaxation factor 
over a time increment $\Delta t$ is
\begin{equation}
\phi_r \equiv \left(1 - e^{-\Delta t / \tau_m}\right).
\end{equation}
In each component, the difference in stress from the mean can be expressed as
a strain increment $\Delta\epsilon_i$ toward equilibrium,
\begin{equation}
\Delta\epsilon_i = \phi_r \omega_i (\tau_i - \bar\tau)
   \left[f_i S(s_i)\right]^{-1}
\end{equation}
where $S$ is the stiffness matrix for the component's material in its
instantaneous state,
\begin{equation}
[S_i]_{kj} \equiv \pdiffl{[\tau_i]_k}{[\epsilon]_j}.
\end{equation}
The strain increment can be expressed as a strain rate,
\begin{equation}
\grad u = \frac{\Delta\epsilon_i}{\Delta t},
\end{equation}
and the change in volume fractions is
\begin{equation}
\pdiffl{f_i}t = \frac 13\Tr\left(\pdiffl{\epsilon_i}t\right).
\end{equation}

In the case of a material described by a scalar EOS, the equilibration
relations can be expressed in simpler form.
Now equilibration acts to reduce differences in the component pressures
$p_i$, functions of the instantaneous states $s_i$.
In each component, the difference in pressure from the mean can be expressed as
a volume change $\Delta f_i$ toward equilibrium,
\begin{equation}
\Delta f_i = \frac 13\Tr\,\omega_i
   (p_i - \bar p)\left(\pdiffl{p_i}{f_i}\right)^{-1}
\end{equation}
where the mean pressure $\bar p$ is the volume-weighted average of the
component pressures.
The isotropic stiffness is
\begin{equation}
\pdiffl{p_i}{f_i} = \dfrac{c_i^2(s_i)\rho_i}{f_i}
\end{equation}
where $c_i$ is the bulk sound speed.

\section{Spall as extension to heterogeneous strength}
\begin{figure}
\begin{center}\includegraphics[scale=0.7]{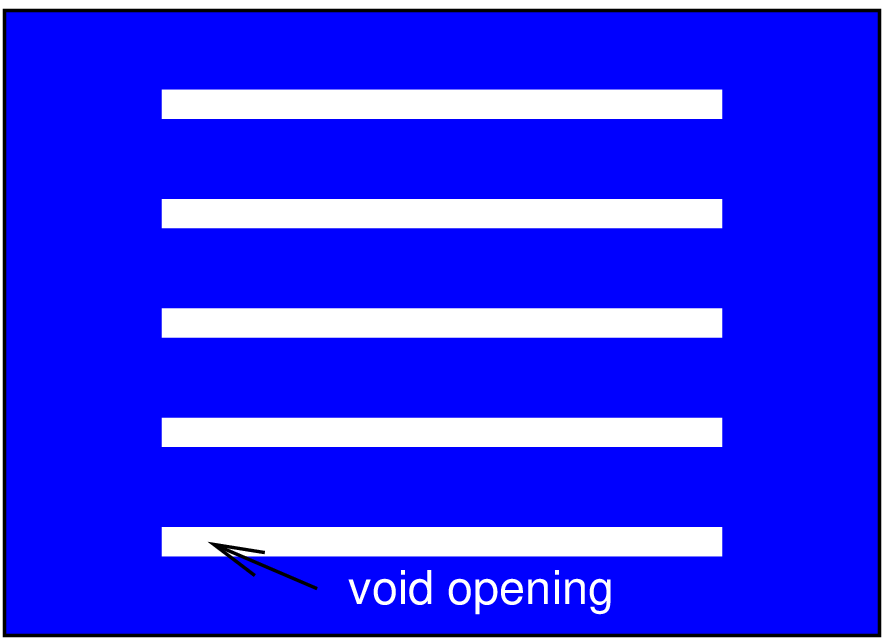}\hfill
\includegraphics[scale=0.7]{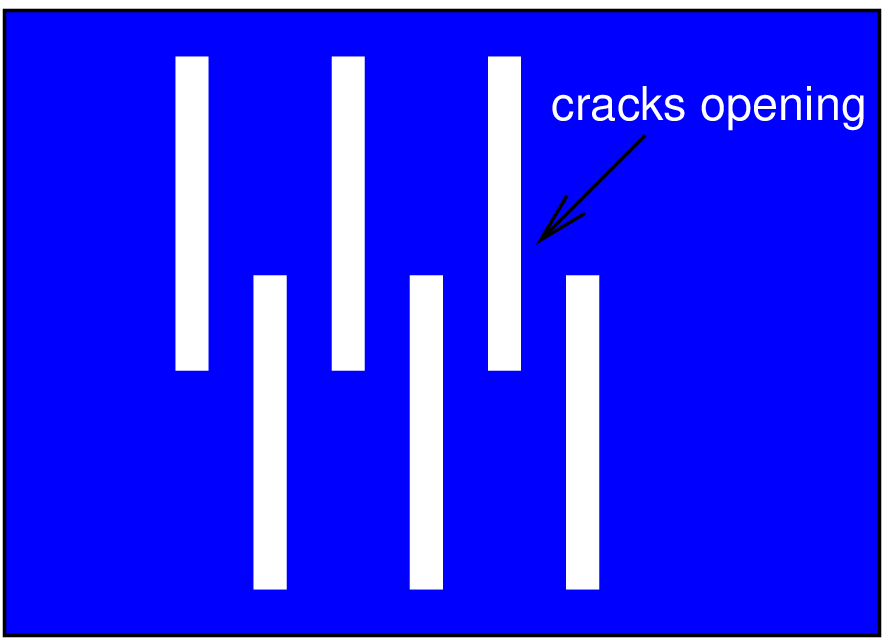}\end{center}
\caption{Schematic of different types of void growth as represented by the
   overlap model, under strain in the horizontal direction:
   ductile (left) and brittle (right).}
\label{fig:poreopen}
\end{figure}

The heterogeneous strength model described above is a natural framework
for developing a class of models for material failure, including spall.
Key elements are the inclusion of void as a component in the mixture,
and the use of the overlap tensor to represent the orientation of growing
regions of void with respect to the direction of maximum tension.
The overlap tensor allows the model to discriminate between ductile and brittle
failure (Fig.~\ref{fig:poreopen}).

The formation and propagation of cracks and voids is dominated by
stress concentrations, i.e. localization phenomena caused by the 
heterogeneity of the microstructure.
Stress concentration is not treated adequately by the overlap tensor,
so sub-scale physics is required, at least at the level of an effective
flow stress for the solid microstructure under tension.
This is similar to the situation with heating induced by pore collapse
in the previous work on heterogeneous high explosives \citep{Mulford01},
where additional shape functions were used to model the enhanced
heating around a collapsing pore, i.e. representing the width of the
temperature distribution as well as the mean.
The heterogeneous model should not be expected to predict the explicit
formation and evolution of pre-existing or incipient damage sites.


The model seamlessly treats recompaction of a damaged or spalled layer,
but may exhibit unphysical healing, unless the model of texture evolution is
sophisticated enough.

\section{Conclusions}
Elastic-plastic data were obtained for Be $(0001)$ crystals using
laser-driven ablative loading, with diagnostics comprising laser Doppler 
velocimetry and {\it in-situ} x-ray diffraction.
These experiments explored the response of Be to dynamic loading on 
nanosecond scales.
The flow stress was much higher than has been observed on microsecond scales,
and exhibited relatively complicated temporal structure.
Important aspects of the velocity history were reproduced using a model
of plastic flow which was based on microstructural processes,
forming the basis of a complete model of crystal plasticity in Be.
A heterogeneous mixture model developed previously for reactive flow studies
was extended to treat strength with a representation of material texture,
and seems to be a promising platform to treat spall.
Accurate models of plasticity and phase changes in the homogeneous components
of the mixture are needed to support physics-based damage and spall models.
Such physics-based models are necessary to relate the large pool of 
experimental data and theoretical work from microsecond time scales and slower
to the qualitatively and quantitatively different behavior observed in
high energy density experiments.

\section*{Acknowledgments}
Scientists contributing to aspects of the experiments and theory include
Dan Thoma and Jason Cooley (group MST-6), Doran Greening (X-7),
Roger Kopp (X-1), Paul Bradley and Doug Wilson (X-2),
Ken McClellan and Darrin Byler (MST-8),
Marcus Knudson (Sandia National Laboratories),
Pedro Peralta and Eric Loomis (Arizona State University).
TRIDENT and P-24 support staff, including
Sam Letzring, Randy Johnson, Bob Gibson, Tom Hurry,  Fred Archuleta, 
Tom Ortiz, Nathan Okamoto, Bernie Carpenter, Scott Evans, and Tom Sedillo,
were instrumental in the laser experiments.
Target fabrication and characterization were carried out by
Ron Perea, Bob Day, Art Nobile, Bob Springer (MST-7), and John Bingert (MST-6).
Funding and project support was provided by
Allan Hauer, Nels Hoffman, Cris Barnes, Steve Batha (Thermonuclear Experiments),
and Aaron Koskelo (Laboratory-Directed Research and Development project
on shock propagation at the mesoscale).

\clearpage
\begin{figure}
\begin{center}
\includegraphics[scale=1.0]{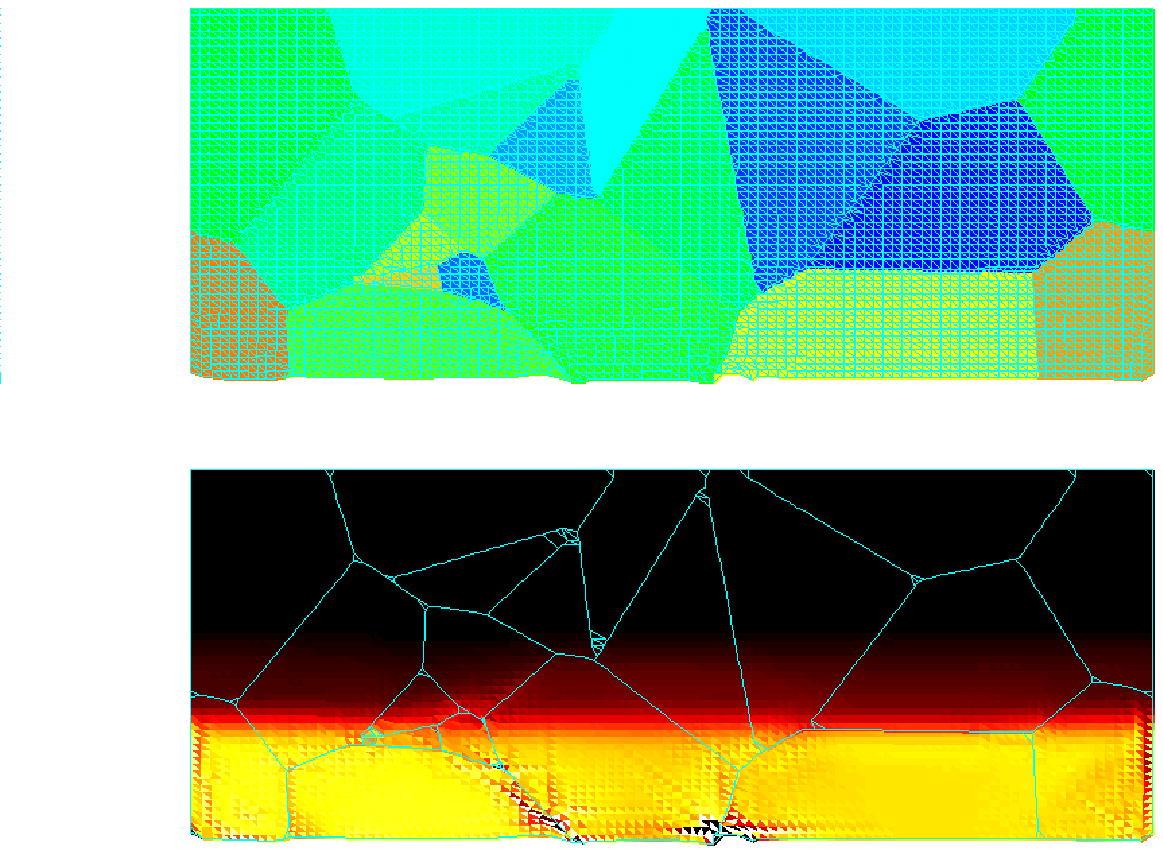}
\end{center}
\caption{Frame from illustrative simulation of shock propagating through
   resolved microstructure in Be.
   Upper figure shows grains (colored according to the crystal orientation)
   and the computational mesh.
   Lower figure shows thermal colormap of mean pressure
   (black: zero, white: 25\,GPa).
   The computational domain is $100\times 50$\,$\mu$m,
   driven by a constant 20\,GPa from the lower surface.
   An elastic precursor (red) can be seen running ahead of the main
   shock, and spatial variations are evident across the width.
   Velocity spectra can be extracted from these simulations and used to
   predict the seeding of implosion instabilities.}
\label{fig:microsim}
\end{figure}

\end{document}